\documentclass{article}
\usepackage{ijcai24}

\usepackage{times}
\usepackage{soul}
\usepackage{url}
\usepackage[hidelinks]{hyperref}
\usepackage[utf8]{inputenc}
\usepackage[small]{caption}
\usepackage{graphicx}
\usepackage{amsmath}
\usepackage{amsthm}
\usepackage{listings}
\usepackage{xcolor}
\usepackage{booktabs}
\usepackage{algorithm}
\usepackage{algorithmic}
\usepackage[switch]{lineno}
\usepackage{subcaption}

\definecolor{codegreen}{rgb}{0,0.6,0}
\definecolor{codegray}{rgb}{0.5,0.5,0.5}
\definecolor{codepurple}{rgb}{0.58,0,0.82}
\definecolor{backcolour}{rgb}{0.95,0.95,0.92}

\lstdefinestyle{mystyle}{
    backgroundcolor=\color{backcolour},   
    commentstyle=\color{codegreen},
    keywordstyle=\color{magenta},
    numberstyle=\tiny\color{codegray},
    stringstyle=\color{codepurple},
    basicstyle=\ttfamily\scriptsize,   %\footnotesize,
    breakatwhitespace=false,         
    breaklines=true,                 
    captionpos=b,                    
    keepspaces=true,                 
    numbers=left,                    
    numbersep=5pt,                  
    showspaces=false,                
    showstringspaces=false,
    showtabs=false,                  
    tabsize=2
}
\title{Supercharging Federated Learning with Flower and NVIDIA FLARE}

\author{
Holger R. Roth$^1$\and
Daniel J. Beutel$^2$\and
Yan Cheng$^1$\and
Javier Fernandez Marques$^2$\and\\
Heng Pan$^2$\and
Chester Chen$^1$\and
Zhihong Zhang$^1$\and
Yuhong Wen$^1$\and
Sean Yang$^1$\and\\
Isaac (Te-Chung) Yang$^1$\and
Yuan-Ting Hsieh$^1$\and
Ziyue Xu$^1$\and
Daguang Xu$^1$\and\\
Nicholas D. Lane$^2$\and
Andrew Feng$^1$
\affiliations
$^1$NVIDIA Corporation\\
$^2$Flower Labs\\
}

\begin{document}

\maketitle

\begin{abstract}
Several open-source systems, such as Flower and NVIDIA FLARE, have been developed in recent years while focusing on different aspects of federated learning (FL). 
Flower is dedicated to implementing a cohesive approach to FL, analytics, and evaluation. Over time, Flower has cultivated extensive strategies and algorithms tailored for FL application development, fostering a vibrant FL community in research and industry.
Conversely, FLARE has prioritized the creation of an enterprise-ready, resilient runtime environment explicitly designed for FL applications in production environments. 
In this paper, we describe our initial integration of both frameworks and show how they can work together to supercharge the FL ecosystem as a whole. Through the seamless integration of Flower and FLARE, applications crafted within the Flower framework can effortlessly operate within the FLARE runtime environment without necessitating any modifications. 
This initial integration streamlines the process, eliminating complexities and ensuring smooth interoperability between the two platforms, thus enhancing the overall efficiency and accessibility of FL applications.

\end{abstract}

\paragraph{Keywords} Federated learning, Open-source, Framework interoperability, Production deployment

\section{Introduction}
%%%%%%%%%%%%%%%%%%%%%%%%%%%%%%%%%%%%%%%%%%%%%%%%%%%%%%%%%%%%%
\begin{figure*}[htbp]
    \centering
    \includegraphics[width=0.75\textwidth]{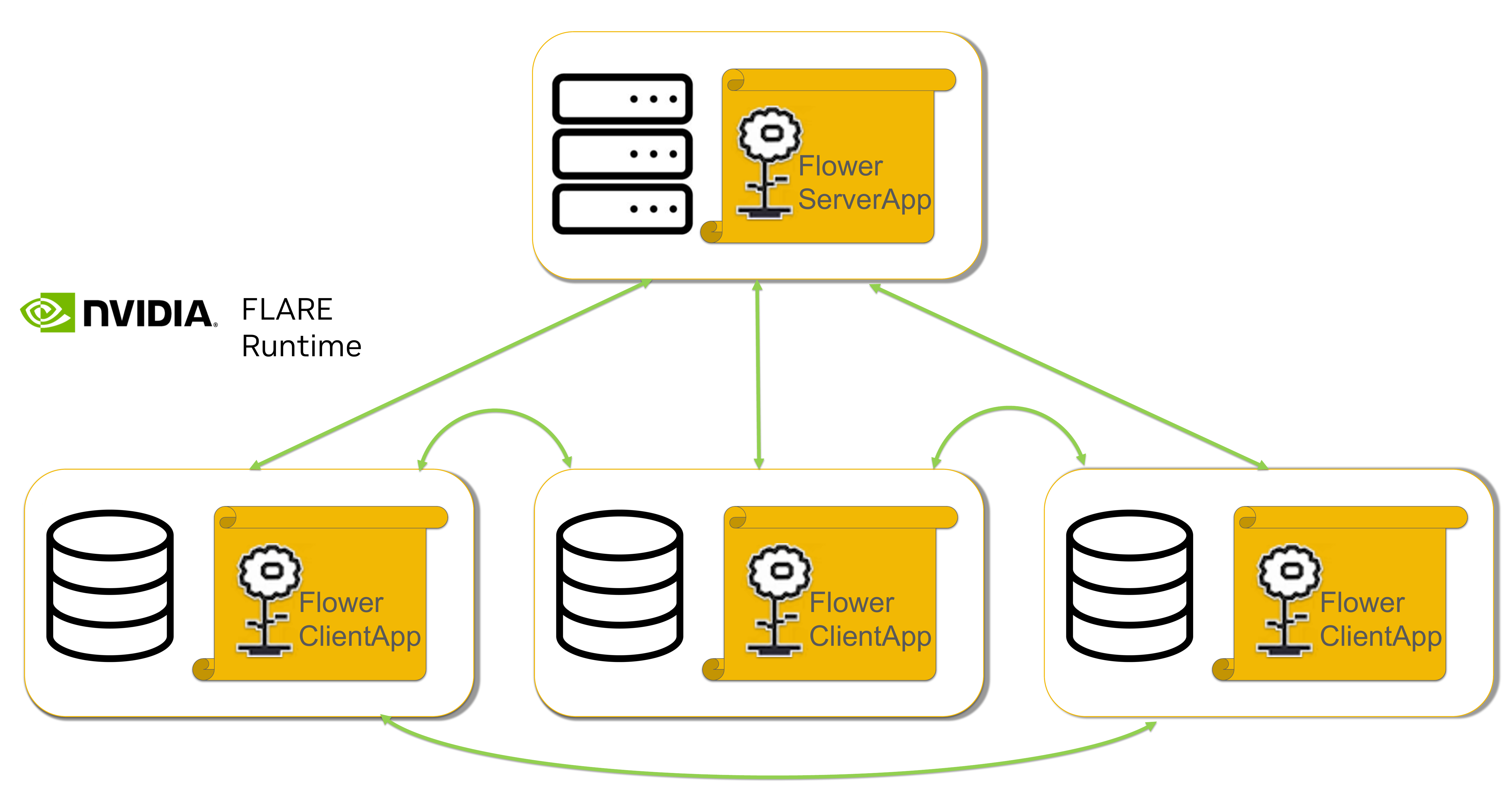}
    \caption{Flower \& NVIDIA FLARE integration. \label{fig:flower_nvflare}}
\end{figure*}
%%%%%%%%%%%%%%%%%%%%%%%%%%%%%%%%%%%%%%%%%%%%%%%%%%%%%%%%%%%%%

Federated learning (FL) makes it possible for AI algorithms to gain experience from a vast range of data located at different sites~\cite{yang2019federated,zhang2021survey,mammen2021federated}. The approach enables several organizations to collaborate on the development of models without directly sharing sensitive data. Preserving privacy while still benefiting from diverse datasets across various locations worldwide.

In this paper, we introduce the collaboration between two of the most widely used solutions for FL, NVIDIA FLARE~\cite{roth2022nvidia} and Flower~\cite{beutel2020flower} to establish compatibility between the two frameworks. With this collaboration, NVIDIA will support running Flower projects on FLARE, thus enabling FLARE developers to access the rich Flower ecosystem, and give Flower developers new options for production deployment through FLARE. The developer communities of both frameworks will be able to build larger-scale consortiums that leverage more data toward training stronger AI models. 
%Flower is a member of NVIDIA Inception, a program designed to help startups of all stages accelerate growth and innovation.

Flower is known for its ease of use, mobile device support, and a large, active open-source community of AI developers and researchers that keep Flower at the forefront of new methods. NVIDIA FLARE offers a comprehensive range of FL features, including robust communication, concurrent job scheduling, security, and confidential computing with strong support for industry-leading NVIDIA hardware. Compatibility between the two complementary frameworks will allow developers to combine the strengths of each platform; under this collaboration, users will be able to build varieties of federated AI systems that were not previously possible. 

Flower \& NVIDIA FLARE integration comes with benefits for both sides. The benefits for FLARE users include using FL algorithms and datasets directly from Flower -- reducing the cost of applying new FL algorithms, leveraging rich built-in differential privacy and secure aggregation support, along with the option to mix FLARE APIs or Flower APIs as required. For Flower users, the benefits include a robust communication framework, concurrent jobs, confidential computing on CPU/GPU, and flexible communication patterns such as server-centric or peer-to-peer. 

We introduced this initiative between Flower and NVIDIA at the Flower AI Summit 2024~\cite{announcing2024nvidia_flower}. This event has grown to be the world’s largest conference on FL. During the opening AI Industry Day keynote, a live demonstration was provided to the audience of the first working prototype to result from this collaboration. This presentation showed FLARE running an unmodified Flower project that performs FL and federated evaluation over several training rounds. This feature is planned to be released to the public with an upcoming FLARE release. We expect that this initial release will be the first of a series of integrations that provide more comprehensive forms of compatibility between Flower and NVIDIA FLARE in the future.

%Flower\footnote{\url{https://flower.ai}} is an open-source project implementing a unified approach to FL, analytics, and evaluation. Flower has developed many strategies and algorithms for FL application development and a healthy FL research community. 

%FLARE\footnote{\url{https://github.com/NVIDIA/FLARE}}, on the other hand, has been focusing on providing an enterprise-ready, robust runtime environment for FL applications. 

%With the integration of Flower and FLARE, applications developed with the Flower framework will run seamlessly in FLARE runtime without the user needing to make any changes. 

%All the user needs to do is configure the Flower application into a FLARE job and submit the job to the FLARE system.

\section{Integration Goals}
Our main goal is to enable users to directly deploy Flower \texttt{ServerApps} and \texttt{ClientsApps} within the FLARE runtime environment. No code changes will be necessary.

Architecturally, Flower uses client/server communication. Flower clients communicate with the server via gRPC\footnote{\url{https://grpc.io}}. FLARE uses the same architecture and allows multiple jobs to run at the same time (multiple jobs share the same set of clients/servers) without requiring multiple ports to be open on the server host.

Since both frameworks follow the same communication architecture, it is relatively easy to make a Flower application a FLARE job by using FLARE as the communicator for the Flower App, as illustrated in Fig.~\ref{fig:flower_nvflare}.

In this proposed approach, Flower \textit{SuperNodes} shift away from direct interaction with the Flower \textit{SuperLink}, opting instead for communication through FLARE. This integration offers several distinctive advantages. Firstly, it facilitates the provisioning of startup kits, including certificates, streamlining the setup process. Moreover, it enables the deployment of custom code, allowing for tailored applications to be implemented seamlessly. User authentication and authorization mechanisms enhance security and access control within the system. Additionally, the inclusion of a \href{https://nvflare.readthedocs.io/en/2.4/user_guide/federated_xgboost/timeout.html}{\texttt{ReliableMessage}} mechanism addresses connection stability concerns, ensuring robust communication. Multiple communication schemes, such as gRPC, HTTP, TCP, Redis, among others, offer flexibility in communication protocols. Notably, FLARE supports peer-to-peer (P2P) communication, which allows for diverse topologies, enabling direct interaction between any parties involved. A multi-job system further enhances efficiency by enabling multiple Flower apps to operate simultaneously without necessitating additional ports on the server. 

\section{Design Principles}
The main design principles of FLARE and Flower share some commonalities, which are key to supporting compatibility and are beneficial for the integration of the two frameworks. We detail them below.

\subsection{FLARE Multi-Job Architecture}
\label{sec:flare_multijob}
%%%%%%%%%%%%%%%%%%%%%%%%%%%%%%%%%%%%%%%%%%%%%%%%%%%%%%%%%%%%%
\begin{figure}[htbp]
    \centering
    \includegraphics[width=1.0\columnwidth]{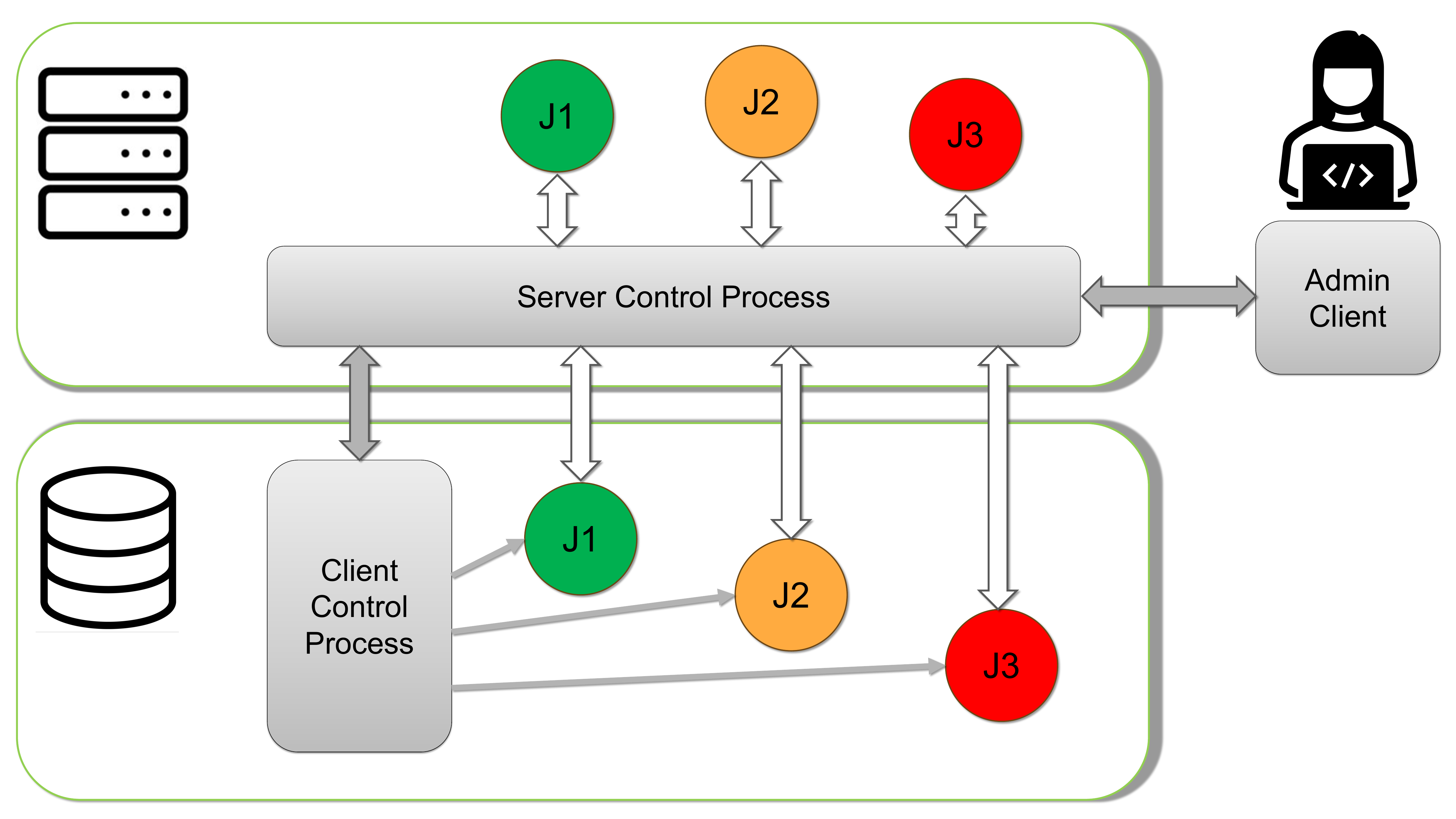}
    \caption{NVIDIA FLARE Multi-job system architecture. \label{fig:multi_job}}
\end{figure}
%%%%%%%%%%%%%%%%%%%%%%%%%%%%%%%%%%%%%%%%%%%%%%%%%%%%%%%%%%%%%

To maximize the utilization of compute resources, FLARE supports multiple jobs running simultaneously, each an independent FL experiment.

As shown in Fig.~\ref{fig:multi_job}, there is the \texttt{Server Control Process} (SCP) on the Server host, and there is a \texttt{Client Control Process} (CCP) on each client host. The SCP communicates with CCPs to manage FLARE jobs (schedule, deploy, monitor, and abort jobs). When the SCP schedules a job, the job is sent to the CCPs of all sites, which create separate processes for the job. These processes form a ``Job Network'' for the job. This network only exists during the runtime of the job. Figure~\ref{fig:multi_job} shows jobs (J1, J2, J3) in different colors on server and client(s). For example, all J1 processes form the Job Network for Job 1.

By default, processes of the same Job Network are not connected directly. Instead, they only connect to the SCP, and all messages between job processes are relayed through the SCP. However, if network policy permits, direct connections could be established automatically between the job processes to obtain maximum communication speed. The underlying communication path is transparent to applications and only requires configuration changes to enable direct communication.

\subsection{Flower Next}
%%%%%%%%%%%%%%%%%%%%%%%%%%%%%%%%%%%%%%%%%%%%%%%%%%%%%%%%%%%%%
\begin{figure}[htbp]
    \centering
    \includegraphics[width=\columnwidth]{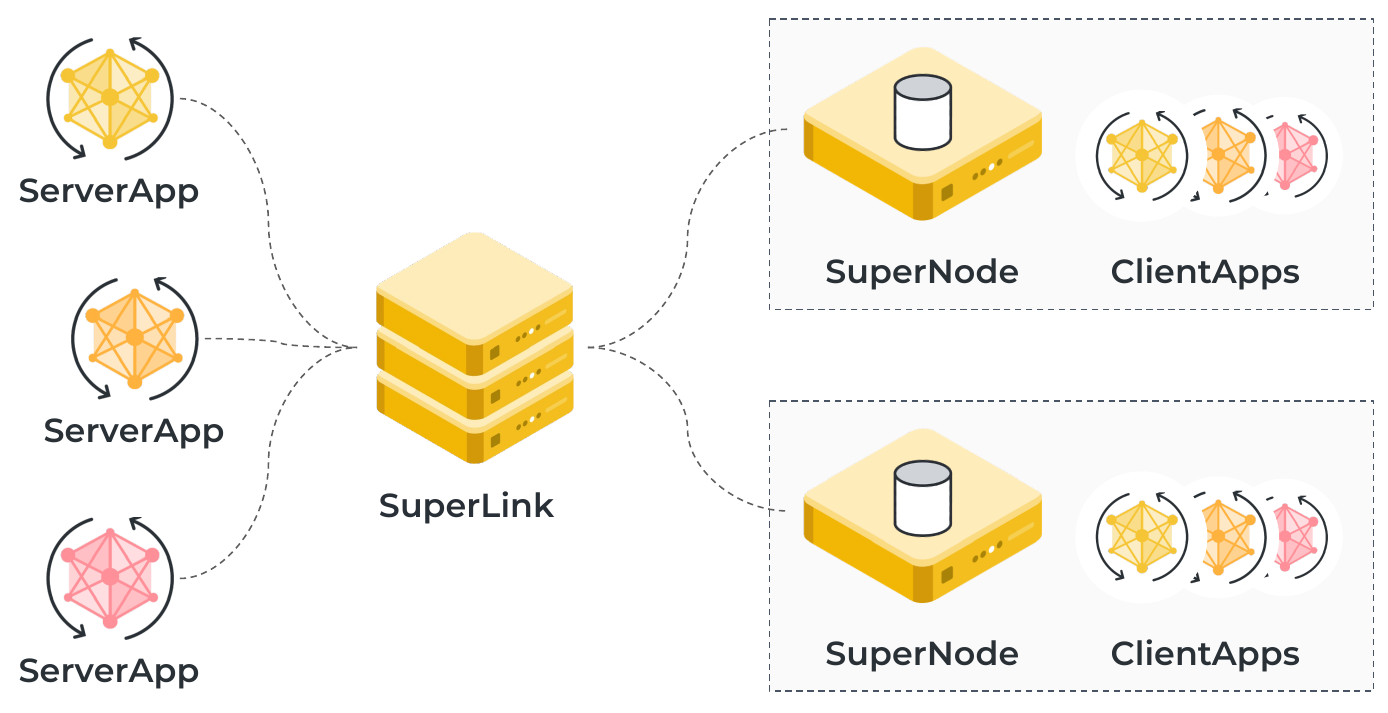}
    \caption{Flower \texttt{SuperLink} and \texttt{SuperNodes}.  \label{fig:flower_next}}
\end{figure}
%%%%%%%%%%%%%%%%%%%%%%%%%%%%%%%%%%%%%%%%%%%%%%%%%%%%%%%%%%%%%
With the recent introduction of Flower Next\footnote{Flower Next is Flower's new high-level API. All new features (like Flower Mods) will be built on top of it.}, Flower enabled multi-run support utilizing long-running server and clients called \texttt{SuperLink} and \texttt{SuperNodes}, respectively, that allow running multiple Flower Apps as illustrated in Fig.~\ref{fig:flower_next}. The \texttt{SuperLink} and \texttt{SuperNode} decouples the communication layer from the Flower \texttt{ServerApp} and \texttt{ClientApps}. This design is very analogous to FLARE's multi-job architecture (see Section~\ref{sec:flare_multijob}) and allows seamless integration, as discussed in the next section.

\section{Initial Integration}
In this first integration, Flower algorithms will be able to utilize the FLARE's reliable messaging system detailed below and other benefits such as metric streaming for experiment tracking.

\subsection{FLARE Reliable Messaging}

The interaction between the FLARE Clients and Server is through reliable messaging. 
First, the requester tries to send the request to the peer. If it fails to send it, it will retry a moment later. This process keeps repeating until the request is sent successfully or the amount of time has passed (which will cause the job to abort).

Secondly, once the request is sent, the requester waits for the response. Once the peer finishes processing, the result is sent to the requester immediately (which could be successful or unsuccessful). At the same time, the requester repeatedly sends queries to get the result from the peer until the result is received or the maximum amount of time has passed (which will cause the job to abort). The result could be received in one of the following ways:

\begin{enumerate}
    \item The result is received from the response message sent by the peer when it finishes the processing.
    \item The result is received from the response to the query message of the requester.
\end{enumerate}

%For details of ReliableMessage, see \url{https://FLARE.readthedocs.io/en/2.4/user_guide/federated_xgboost/timeout.html}

\subsection{Integration Design}
%%%%%%%%%%%%%%%%%%%%%%%%%%%%%%%%%%%%%%%%%%%%%%%%%%%%%%%%%%%%%
\begin{figure}[htbp]
    \centering
    \includegraphics[width=\columnwidth]{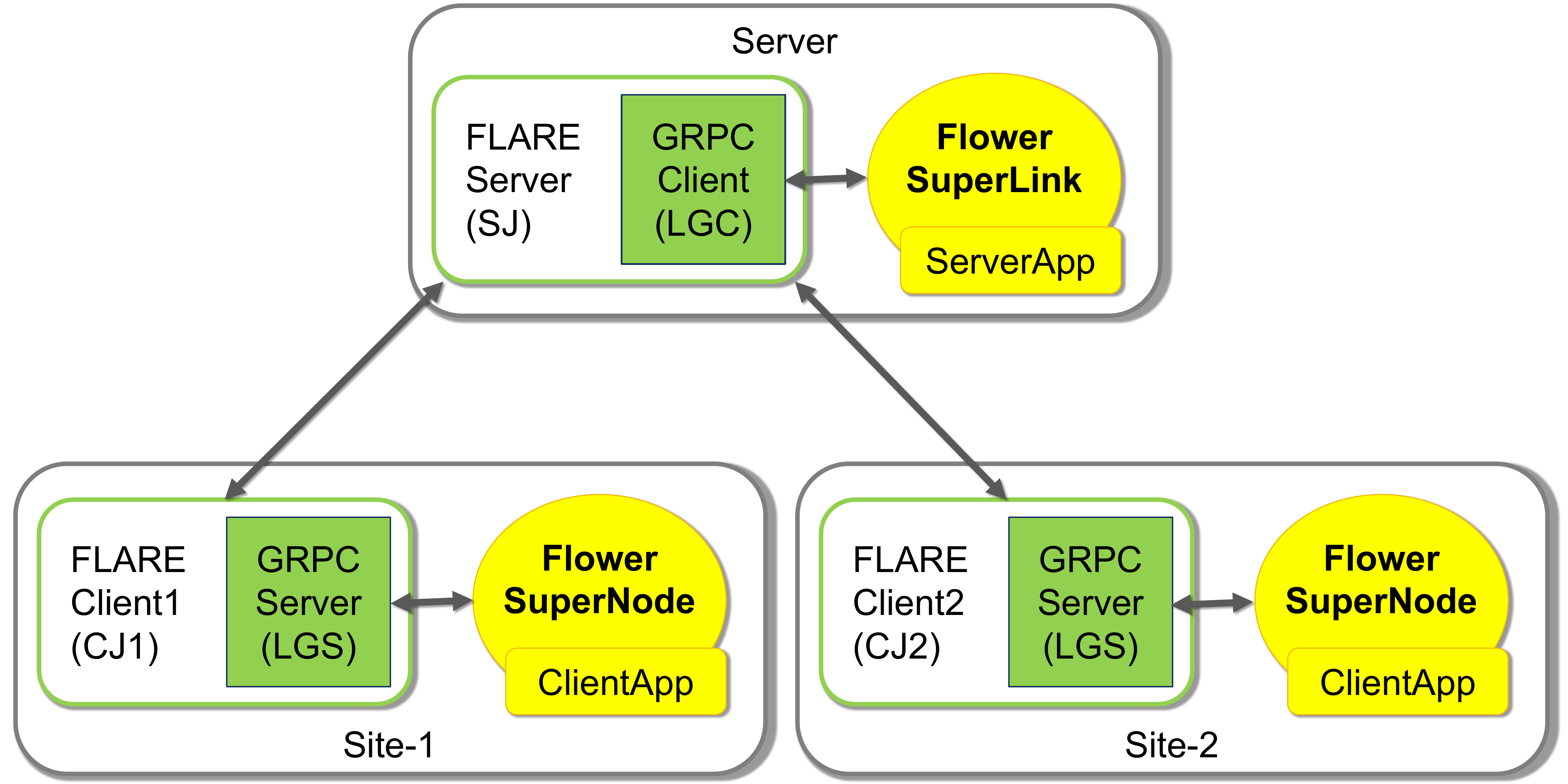}
    \caption{Integration of Flower Apps within FLARE. \label{fig:flower_integration}}
\end{figure}
%%%%%%%%%%%%%%%%%%%%%%%%%%%%%%%%%%%%%%%%%%%%%%%%%%%%%%%%%%%%%

Flower uses gRPC as the communication protocol. To use FLARE  as the communicator, we route Flower’s gRPC messages through FLARE. To do so, we change the server endpoint of each Flower client to a local gRPC server (LGS) within the FLARE client.  

As shown in Fig.~\ref{fig:flower_integration}, there is a \texttt{Local GRPC server} (LGS) for each site that serves as the server endpoint for the Flower \texttt{SuperNode} on the site. Similarly, there is a \texttt{Local GRPC Client} (LGC) on the FLARE Server that interacts with the Flower \texttt{SuperLink}. The message path between the Flower \texttt{SuperNode} and the Flower \texttt{SuperLink} is as follows:

\begin{enumerate}
    \item The Flower \texttt{SuperNode} generates a gRPC message and sends it to the LGS in the FLARE Client
    \item FLARE Client forwards the message to the FLARE Server -- This is a reliable FLARE message.
    \item FLARE Server uses the LGC to send the message to the Flower \texttt{SuperLink}.
    \item The Flower \texttt{SuperLink} sends the response back to the LGC in the FLARE Server.
    \item FLARE Server sends the response back to the FLARE Client.
    \item FLARE Client sends the response back to the Flower \texttt{SuperNode} via the LGS.
\end{enumerate}
Note that the Flower \texttt{SuperNode} could be running as a separate process or within the same process as the FLARE Client.

This integration design enables users to directly deploy Flower ServerApps and ClientsApps within the FLARE Runtime Environment without requiring any code changes.

\section{Experiments}
Next, we describe two experiments. First, an integration of running Flower Apps in FLARE's environment without any code changes needed, followed by a tighter integration, allowing a Flower App to utilize some of FLARE's communication features, such as the metric streaming during training used for experiment tracking.

\subsection{Integration Without Code Changes}
To showcase our integration, we run Flower's PyTorch-Quickstart example directly within FLARE. The Flower creation of Flower ServerApp and ClientApps in shown in Listings~\ref{lst:flower_server}~and~\ref{lst:flower_client}. 

\begin{lstlisting}[language=Python, caption=Creating a Flower ServerApp., label=lst:flower_server]
# Define strategy
strategy = FedAdam(...)

# Create Flower ServerApp
app = ServerApp(
    config=ServerConfig(num_rounds=3),
    strategy=strategy,
)
\end{lstlisting}

\begin{lstlisting}[language=Python, caption=Creating a Flower ClientApp., label=lst:flower_client]
class FlowerClient(NumPyClient):
  def fit(self, parameters, config):
    model.set_weights(parameters)
    model.fit(x_train, y_train, epochs=1, batch_size=32)
    return model.get_weights(), len(x_train), {}

  def evaluate(self, parameters, config):
    model.set_weights(parameters)
    loss, accuracy = model.evaluate(x_test, y_test)
    return loss, len(x_test), {"accuracy": accuracy}


def client_fn(cid: str):
    """Return an instance of Flower `Client`."""
    return FlowerClient().to_client()


# Create Flower ClientApp
app = ClientApp(
    client_fn=client_fn,
)
\end{lstlisting}
With the integration of Flower and FLARE, applications developed with the Flower framework will run seamlessly in FLARE runtime without the user needing to make any changes. 

%All the user needs to do is configure the Flower application into a FLARE job and submit the job to the FLARE system.

We have two options to deploy a Flower project within the FLARE runtime. 
\begin{enumerate}
    \item Use FLARE's simulator or CLI:\\ \texttt{nvflare job submit <job\_path>}
    \item Use Flower Run CLI:\\ \texttt{flwr run <project\_path>}
\end{enumerate}

\begin{figure*}[!tbp]
     \centering
     \begin{subfigure}[b]{\columnwidth}
         \centering
         \includegraphics[width=0.9\textwidth]{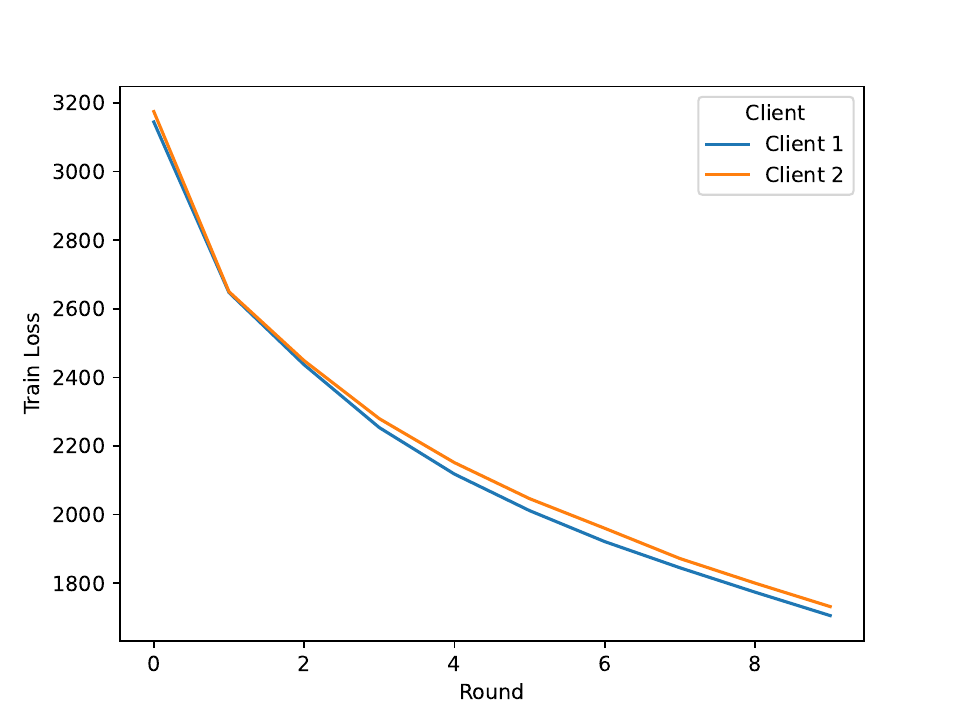}
         \caption{Flower alone.}
         \label{fig:repr_flwr_alone}
     \end{subfigure}
     \hfill
     \begin{subfigure}[b]{\columnwidth}
         \centering
         \includegraphics[width=0.9\textwidth]{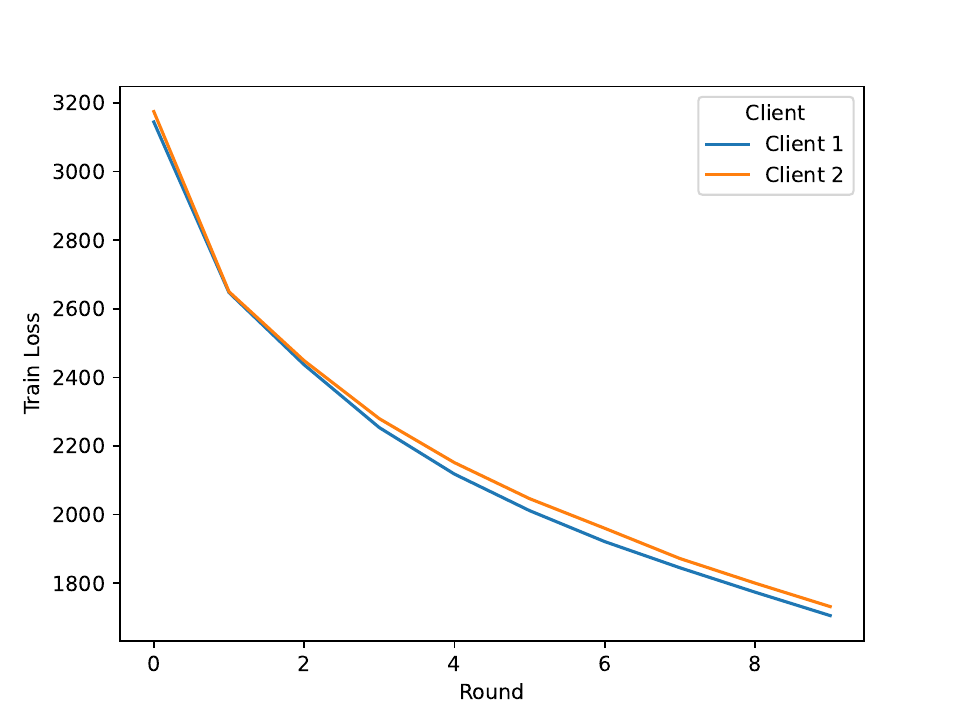}
         \caption{Flower in FLARE.}
         \label{fig:repr_flwr_in_flare}
     \end{subfigure}
     \hfill
        \caption{Comparison of running a Flower application natively (a) or within FLARE (b).}
        \label{fig:reproduce}
\end{figure*}

\paragraph{Reproducibility:} Here, we show that one gets the same results when running a Flower application alone compared to running it within the FLARE environment. We initialize the code using the same random seeds for the clients and plot training curves in Fig.~\ref{fig:reproduce}. Both graphs will match exactly when overlaid, showing that the messages routed by FLARE do not influence the results.
\subsection{Hybrid Integration Using FLARE's Experiment Tracking}
By deploying a Flower App within the FLARE runtime, we can now utilize features from both frameworks. For example, we can utilize FLARE's experiment tracking feature within the Flower client code as shown in Listing~\ref{lst:experiment}.
\begin{lstlisting}[language=Python, caption=Usage of FLARE's experiment tracking within a Flower ClientApp., label=lst:experiment]
from nvflare.client.tracking import SummaryWriter
writer = SummaryWriter()

def train(net, trainloader, epochs):
    """Train the model on the training set."""
    criterion = torch.nn.CrossEntropyLoss()
    optimizer = torch.optim.SGD(net.parameters(), lr=0.001, momentum=0.9)

    global TRAIN_STEP
    for _ in range(epochs):
        avg_loss = 0.0
        for batch in tqdm(trainloader, "Training"):
            images = batch["img"]
            labels = batch["label"]
            optimizer.zero_grad()
            loss = criterion(net(images.to(DEVICE)), labels.to(DEVICE))
            loss.backward()
            optimizer.step()
            avg_loss += loss.item()
        writer.add_scalar("train_loss", avg_loss/len(trainloader), TRAIN_STEP)
        TRAIN_STEP += 1
...

\end{lstlisting}

\begin{figure}[htbp]
    \centering
    \includegraphics[width=1.0\columnwidth]{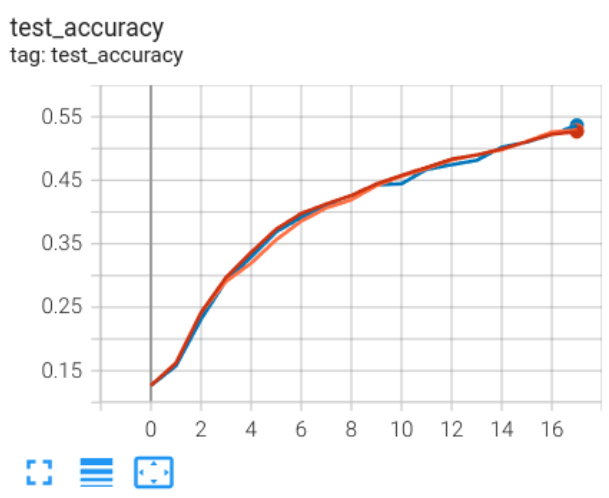}
    \caption{Flower ClientApps running with FLARE's experiment tracking and visualized using TensorBoard. \label{fig:exp_tracking}}
\end{figure}

Once the experiment runs, we can see the experiment tracking metrics from each client being streamed to the FLARE server as shown in~Fig.~\ref{fig:exp_tracking} in the case of three clients.
%%%%%%%%%%%%%%%%%%%%%%%%%%%%%%%%%%%%%%%%%%%%%%%%%%%%%%%%%%%%%
%%%%%%%%%%%%%%%%%%%%%%%%%%%%%%%%%%%%%%%%%%%%%%%%%%%%%%%%%%%%%

\section{Discussion}
The integration of Flower with NVIDIA FLARE offers mutual benefits to both platforms. For FLARE users, this integration allows for direct utilization of FL algorithms and datasets from Flower, thereby reducing the cost associated with implementing new FL algorithms. Users will also have the flexibility to leverage either FLARE APIs or Flower APIs, providing versatility in their development approach.

On the other hand, Flower users also reap significant advantages from this integration. They gain access to robust communication capabilities, enabling seamless data exchange between Flower and FLARE. Concurrent job execution enhances efficiency by enabling multiple tasks to be processed simultaneously. Furthermore, Flower users benefit from the integration's support for confidential computing on both CPU and GPU, ensuring data security and privacy. The integration also maintains a server-centric architecture while facilitating peer-to-peer communication, offering a comprehensive and versatile solution for FL applications.

Looking ahead, we plan to enable FLARE users to leverage Flower's rich mobile and IoT device support to incorporate these into their federations more seamlessly. Furthermore, the potential for supporting very large messages, up to hundreds of gigabytes, presents an exciting prospect~\cite{roth2024empowering}. This would require integration with \href{https://nvflare.readthedocs.io/en/main/user_guide/configurations/communication_configuration.html}{CellNet} in Flower, promising even greater scalability and capability for handling large-scale data processing tasks, which is crucial for building reliable foundation models and is part of our future collaboration plan.

%% The file named.bst is a bibliography style file for BibTeX 0.99c
\bibliographystyle{named}
\bibliography{ref}

\end{document}